\documentclass{aastex62}

\definecolor{ForrestGreen}{rgb}{0.133,0.545,0.133}

\usepackage{CJK}

\submitjournal{ApJ}

\shorttitle{Coronal Mini-jets}
\shortauthors{Chen et al.}

\begin{document}
\title{Coronal Mini-jets in an Activated Solar Tornado-like Prominence}

\begin{CJK*}{UTF8}{gbsn}
\correspondingauthor{Huadong Chen}
\email{hdchen@nao.cas.cn}
\correspondingauthor{Jun Zhang}
\email{zjun@ahu.edu.cn}

\author[0000-0001-6076-9370]{Huadong Chen (陈华东)}
\affil{CAS Key Laboratory of Solar Activity, 
     National Astronomical Observatories, 
     Chinese Academy of Sciences, 
      Beijing 100101, People's Republic of China}
\affil{School of Astronomy and Space Science, University of Chinese Academy of Sciences, Beijing 100049, People's Republic of China}

\author{Jun Zhang (张军)}
\affil{School of Physics and Materials Science, Anhui University, Hefei 230601, People's Republic of China}
\affil{CAS Key Laboratory of Solar Activity, 
     National Astronomical Observatories, 
     Chinese Academy of Sciences, 
      Beijing 100101, People's Republic of China}


\author{Bart De Pontieu}
\affiliation{Lockheed Martin Solar and Astrophysics Laboratory, Palo Alto, CA 94304, USA}

\author[0000-0002-5431-6065]{Suli Ma (马素丽)}
\affiliation{CAS Key Laboratory of Solar Activity, 
     National Astronomical Observatories, 
     Chinese Academy of Sciences, 
      Beijing 100101, People's Republic of China}

\author[0000-0002-5740-8803]{Bernhard Kliem}
\affiliation{Institute of Physics and Astronomy, University of Potsdam, Potsdam 14476, Germany}

\author{Eric Priest}
\affiliation{School of Mathematics and Statistics, University of St Andrews, St Andrews, Fife KY16 9SS, UK}

\begin{abstract}
High-resolution observations from the $Interface~Region~Imaging~Spectrometer$ ($IRIS$) reveal the existence of a particular type of small solar jets, which arose singly or in clusters from a tornado-like prominence suspended in the corona.
In this study, we perform a detailed statistical analysis of 43 selected mini-jets in the tornado event.
Our results show that the mini-jets typically have: (1) a projected length of 1.0--6.0 Mm, (2) a width of 0.2--1.0 Mm, (3) a lifetime of 10--50 s, (4) a velocity of 100--350 km s$^{-1}$, and (5) an acceleration of 3--20 km s$^{-2}$.
Based on spectral diagnostics and EM-Loci analysis, these jets seem to be multi-thermal small-scale plasma ejections with an estimated average electron density of $\sim$2.4 $\times$ 10$^{10}$ cm$^{-3}$ and an approximate mean temperature of $\sim$2.6 $\times$ 10$^{5}$ K.
Their mean kinetic energy density, thermal energy density and dissipated magnetic field strength are roughly estimated to be $\sim$9 erg cm$^{-3}$, 3 erg cm$^{-3}$, and 16 G, respectively.
The accelerations of the mini-jets, the UV and EUV brightenings at the footpoints of some mini-jets, and the activation of the host prominence suggest that 
the tornado mini-jets are probably created by fine-scale external or internal magnetic reconnections (a) between the prominence field and the enveloping or background field or (b) between twisted or braided flux tubes within the prominence.
The observations provide insight into the geometry of such reconnection events in the corona and have implications for the structure of the prominence magnetic field and the instability that is responsible for the eruption of prominences and coronal mass ejections.
\end{abstract}


\section{Introduction} \label{sec:intro}
Solar jets are transient collimated plasma ejections in the solar atmosphere \citep[][]{roy73}.
They are thought to be ejected along open magnetic fields or the legs of large-scale magnetic loops \citep[e.g.,][]{shibata94a, liuy05}.
As space-borne instruments have evolved since the 1980's, the observations of dynamic solar events have been extended from H$_{\alpha}$ and radio to UV, EUV, and X-ray wavebands \citep[e.g.,][]{schmahl81, schmieder88, alexander99, zhangj00, cirtain07, jiangy07, chenh08, tianh11, joshi18, zhangq19}.
According to relevant studies \citep[e.g.,][]{shimojo96, savcheva07}, large-scale solar jets can extend to lengths of $\sim$10$^5$ km and widths of $\sim$10$^4$ km; they have typical speeds on the order of a few $\times$ 10$^2$ km s$^{-1}$ and lifetimes ranging from several minutes to a few hours.

Allowing for a high degree of correlation between jets and photospheric magnetic flux activity, such as flux emergence and cancellation \citep[e.g.,][]{roy73, golub81, chae99, liuy04, jiangy07, chenh08, yangl11}, many authors have been inclined to believe that jets result from magnetic reconnection between potential or twisted magnetic loops and ambient open fields  \citep[e.g.,][]{heyvaerts77, forbes84, shibata86, canfield96, patsourakos08, kamio10, pariat10, yanglp18, lid19}.
In contrast to this sort of ``standard'' jet, another type termed ``blowout'' jet was proposed by \citet{moore10}, in which jets are associated with eruptions of miniature filaments. 
\citet{sterling15} further found that a minifilament eruption could be found in each of 20 randomly selected X-ray jets formed in polar coronal holes.
Up to the present, a substantial amount of observations \citep[e.g.,][]{hongj11, hongj16, sheny12, sheny17, young14, lee15, lix15, sterling16, zhangy17, kumar18} and numerical simulations \citep[e.g.,][]{archontis13, pariat15, pariat16, wyper18, meyer19} have shown that the blowout eruption of a small-scale sheared-core magnetic arcade can play an important role in producing a solar jet.
It is also worth noting that \citet{lix19} reported some jet-like features, which were rooted in the ribbons of an X-class flare and might be caused by chromospheric evaporation.

Even though magnetic reconnection seems to be necessary for the occurrence of most solar jets, the ways reconnection occurs during jet formation may be remarkably different from each other, thus leading to a diversity of jet morphology.
A multitude of studies have mentioned the spinning motion of jets \citep[e.g.,][]{liuw09, chenh12, hongj13, sheny11, schmieder13, zhangq14, liuj18, lul19}, which is generally considered to be a result of relaxation of magnetic twist through reconnection \citep[e.g.,][]{canfield96, fang14} or the conversion of mutual magnetic helicity into self-helicity during three-dimensional reconnection \citep[][]{priest16}.
A rare event of coronal twin jets was presented by \citet{liuj16}.
\citet{hongj19} found that a solar jet was accompanied by oscillatory reconnection.
\citet{shibata94b} categorized jets as anemone type or two-sided-loop type, which is associated with relatively vertical or horizontal overlying coronal field configurations, respectively.
Recently, \citet{zhengr18} provided an example of a two-sided-loop jet related to ejected plasmoids and twisted overlying fields.
\citet{sterling19}, \citet{sheny19}, and \citet{yangb19} further found that two-sided-loop jets can also be driven by eruptions of mini-filaments below overlying large magnetic loops. 

Besides large EUV or X-ray coronal jets, high-resolution observations have revealed that small-scale jet activity takes place more frequently than large jets \citep[e.g.,][]{depontieu04, shibata07, tianh14a, young18}.
They are ubiquitous in the lower solar atmosphere, such as spicules observed at the limb  \citep[][]{depontieu07}, chromospheric anemone jets outside active regions \citep[][]{shibata07, nishizuka11}, penumbral microjets in sunspots \citep[][]{katsukawa07, Esteban19}, transition region network jets \citep[][]{tianh14a, kayshap18, cheny19}, and intermittent jets from light bridges of sunspots  \citep[][]{houy17, tianh18}. 
Small-scale jets are usually one or two orders of magnitude smaller than large jets and have a shorter life span varying from dozens of seconds to several minutes.
In terms of dynamics, there seem to be two kinds of small jet, which have a speed of $\sim$50 km s$^{-1}$ and $\sim$150 km s$^{-1}$, respectively.
\citet{depontieu07} first proposed that the two types of small jets or spicules dominating the solar chromosphere are separately driven by shock waves (Type-I) and magnetic reconnection (Type-II).
Two similar sorts of small jet were also found from sunspot light bridges by \citet{houy17} and \citet{tianh18}.

Up to now, the triggering mechanism of small jets has not been fully understood.
Many models were devoted to interpreting their formation.
\citet{judge11} suggested that some populations of spicules and fibrils correspond to warps in two-dimensional sheet-like structures.
\citet{takasao13} found that slow-mode shock waves generated by magnetic reconnection in the chromosphere and photosphere play key roles in accelerating chromospheric jets.
\citet{cranmer15} modeled spicules as narrow, intermittent extensions of the chromosphere using the output of a time-dependent simulation of reduced magnetohydrodynamic (MHD) turbulence. 
The MHD simulations performed by \citet{martnez17} and \citet{depontieu17} revealed a novel driving mechanism for spicules in which ambipolar diffusion resulting from ion-neutral interactions plays a dominant role.
\citet{tianh18} studied the fine-scale jets from sunspot light bridges. 
The inverted Y-shape structure of the jets they observed does not seem to be easily explained by non-reconnection models.
Recently, \citet{samanta19} detected flux emergence and/or flux cancellation around the spicule footpoint region and conjectured that this supports the formation of spicules from reconnection.
Their observations do not exclude other formation mechanisms of small jets \citep[e.g.,][]{martnez17}.

Recently, three-dimensional MHD and radiative MHD numerical experiments have shown how flux emergence can drive the formation of jets in the low solar atmosphere.  \citet{raouafi16} gave an excellent overview of observations and models of jets. 
\citet{moreno13} modelled their production in coronal holes, while \citet{moreno18} modelled small-scale flux emergence. \citet{nobrega16} considered cool surges, while \citet{nobrega17} explained observed transition-region properties of surges, and \citet{nobrega20} incorporated nonequilibrium ionisation and ambipolar diffusion.

In this study, we consider a particular type of small-scale jet, which was first mentioned by \citet{chenh17}.
Different from the usual jets previously reported, these small jets did not emanate from the photosphere or chromosphere, but directly appeared in a tornado-like prominence suspended in the corona.
This appears to be a very rare phenomenon.
The formation and disintegration mechanism of such prominences has been investigated by 
\citet{chenh17}.
Here, we focus on statistical information about the dynamical and energetic characteristics of these unusual coronal mini-jets and their possible triggering mechanism.
In the next section, we describe the observational data. This is followed by a detailed statistical investigation of the dynamical and energetic properties of the mini-jets.
Finally, we summarize and discuss the results.

\section{Observations} \label{sec:obser}
On 2015 March 19, the $Interface~Region~Imaging~Spectrometer$ \citep[$IRIS$;][]{depontieu14} slit-jaw imager (SJI) provided the 1330 \AA\ intensity images with a spatial scale of 0\farcs33 and a cadence of 9.3 s.
The $IRIS$ spectral data were taken in a large coarse 8-step raster mode with a 74 s cadence and a spectral resolution of $\sim$0.025 \AA\ in the far ultra-violet (FUV) waveband.
We mainly used the emissions of the \ion{O}{4} line pair (1399.8 \AA\ and 1401.2 \AA) in the vicinity of the \ion{Si}{4} 1402.8 \AA\ line to estimate the electron densities of the jets.
The IRIS data have been summed spatially.
Most of the mini-jets were also captured by the Atmospheric Imaging Assembly \citep[AIA;][]{lemen12} on board the $Solar$ $Dynamics$ $Observatory$ \citep[$SDO$;][]{pesnell12}, which supplies us with full-disk intensity images up to 0.5 R$_{\sun}$ above the solar limb with 0\farcs6 pixel size and 12 s cadence in 7 EUV channels centered at 
304 \AA\ (\ion{He}{2}, 0.05 MK),
131 \AA\ (\ion{Fe}{8}, 0.4 MK and \ion{Fe}{21}, 11 MK),
171 \AA\ (\ion{Fe}{9}, 0.6 MK), 
193 \AA\ (\ion{Fe}{12}, 1.3 MK and \ion{Fe}{24}, 20 MK),
211 \AA\ (\ion{Fe}{14}, 2 MK), 
335\AA\ (\ion{Fe}{16}, 2.5 MK), 
and 94 \AA\ (\ion{Fe}{18}, 7 MK),
respectively.
One longitudinal magnetogram with a 0\farcs5 plate scale from the Helioseismic and Magnetic Imager \citep[HMI;][]{schou12} on board $SDO$ was utilized to show the active region AR 12297 as the background of the magnetic field lines from a potential field source surface extrapolation \citep[PFSS; e.g.,][]{schatten69}.

\section{Results} \label{}
\begin{figure}
\epsscale{1}
\plotone{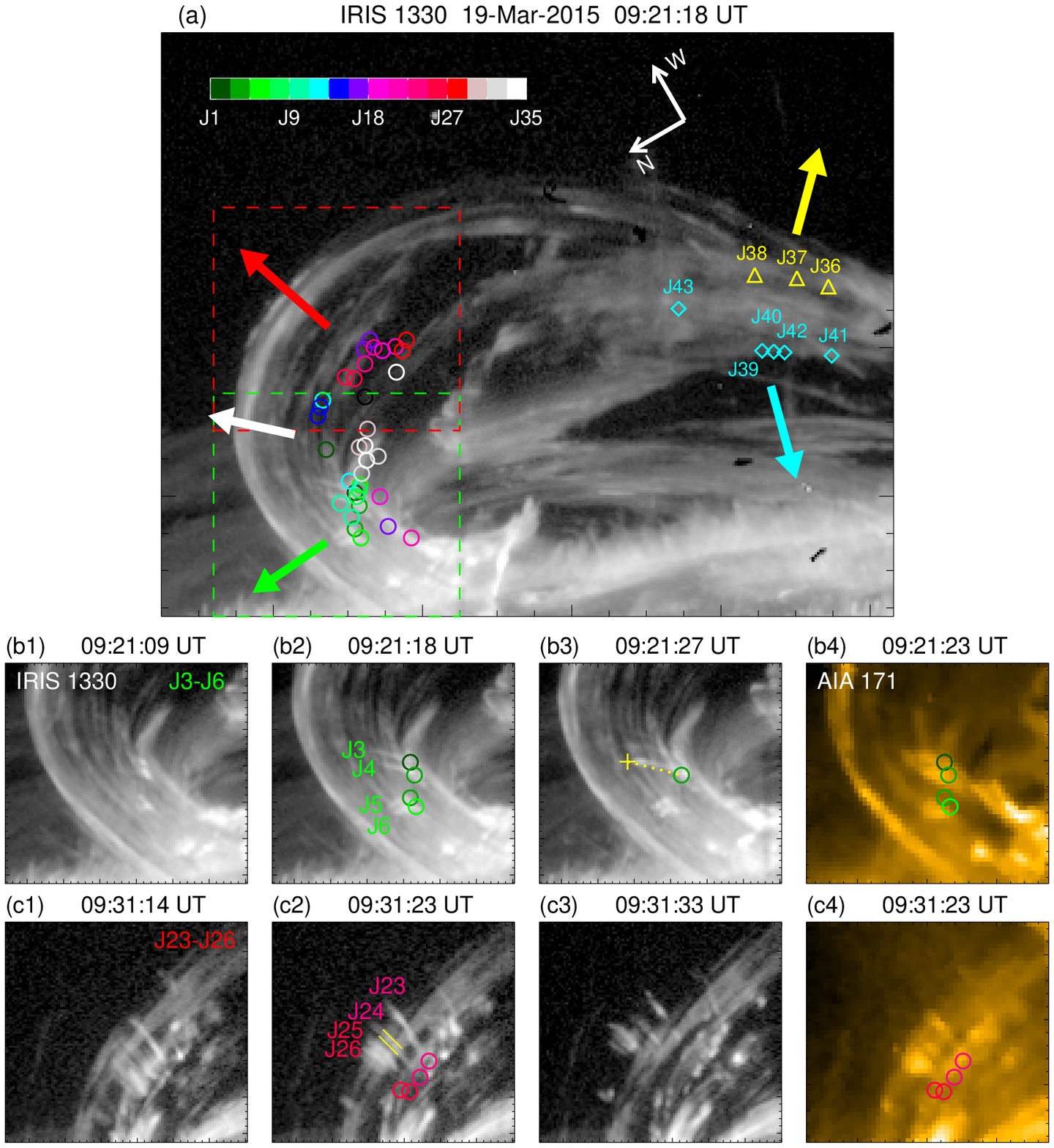}
\caption{(a) The locations of coronal mini-jets (J1--J43) are marked in $IRIS$ 1330 \AA\ SJI images taken at 09:21:18 UT.
The circles, triangles, and diamonds represent the footpoint positions of J1--J35, J36--J38, and J39--J43, respectively.
The thick arrows approximately indicate the ejection directions of the jets.
The green and red boxes in panel (a) separately correspond to the fields of view (FOVs) of panels (b1)--(b4) and (c1)--(c4).
The $IRIS$ 1330 \AA\ SJI images ((b1)--(b3)) and AIA 171 \AA\ image (b4) show the evolutions of J3--J6; ((c1)--(c4)) are the same as ((b1)--(b4)), but for J23--J26.
The plus in panel (b3) denotes the top edge of J4.
The distance between the two short lines in panel (c2) indicates the projected width of J24.
All images have been rotated counterclockwise by 120\degr\ for convenience.
The center of panel (a) is at solar (x,y) = (933\arcsec, --354\arcsec) and 
The FOV is 98\arcsec\ $\times$ 78\arcsec.
An animation of the $IRIS$~1300 \AA\ SJI images is available in the online Journal. The animated images run from 09:10 to 09:50~UT. 
\label{fig1}}
\end{figure}

During 2015 March 19--20, two tornado-like prominences successively formed and developed near active region AR 12297 ($\sim$S16W79). 
In the early evolution process of the first tornado, a multitude of small-scale jet-like structures (mini-jets) seem to be rooted in and were ejected from the thread structures of the activated tornado.
We selected 43 mini-jets (J1--J43) in total, which took place during the period of 09:17--09:40 UT and clearly showed their collimated structures and dynamical evolutions in the high-resolution $IRIS$ 1330 \AA\ SJI images (see the online animated version of Figure~\ref{fig1}).
We marked their footpoint positions with a circle (J1--J35), triangle (J36--J38), and diamond (J39--J43) in the SJI image taken at 09:21:18 UT (Figure~1(a)).
Unlike the flows along the threads of a prominence \citep[e.g.,][]{chenh16}, these jets were expelled approximately perpendicular to the local prominence's axes, as indicated by the arrows in Figure~1(a).
Another remarkable feature is that the jets sometimes appeared in clusters happening almost simultaneously and being very close to each other in space with approximately parallel ejection directions.
The evolutions of several groups of clustered mini-jets are presented in the SJI 1330 \AA\ images in the middle (J3--J6) and bottom (J23--26) panels of Figure~1.
Two AIA 171 \AA\ images are also given in Figure~1(b4) and (c4) to show the eight mini-jets in the EUV line.
It can be seen that the spatial scales of these jets are so small that some of them, such as J5--J6 in the panel (b4) and J25--J26 in the panel (c4), can be hardly distinguished from each other in the 171 \AA\ images.

\subsection{Characteristics in Time, Space and Dynamics} \label{subsec:}
Based on the $IRIS$ 1330 \AA\ SJI data, we characterized 43 mini-jets with a statistical analysis of their temporal and spatial scales and dynamics, including the projected length ($l$), width ($w$), velocity ($v_{j}$), acceleration ($a$), lifetime ($\tau$) etc.
The results are listed in the left columns of Table~1.
The lengths of the jets are defined as the distances between their footpoints and the farthest  top edges as measured in the directions of jet propagation (see the dotted line in Figure~1(b3)).
Assuming that the mini-jets moved along the magnetic flux tubes, it is reasonable to conjecture that they have a cylindrical structure.
We measured their widths at their midpoints, as denoted by the distance between the two short lines in Figure~1(c2).
The jet lifetimes are on the order of tens of seconds, which is not much longer than the temporal resolutions ($\sim$10 s) of the SJI and AIA observations.
Sometimes, it is hard to track the entire evolution of the jets, as they may appear and/or disappear during the gap between two successive intensity images.
We approximately calculated the velocities of the mini-jets by dividing their lengths by the corresponding time lags and further derived the accelerations from the velocities and the time lags under the assumption of a zero initial speed.
Figures~2(a)--(e) present the distributions of the $l$, $w$, $\tau$, $v_{j}$, and $a$, respectively. 
It can be seen that most apparent velocities are less than 350 km s$^{-1}$, while accelerations are typically less than 20 km s$^{-2}$.
The dashed lines in Figures~2(a)--(e) indicate the mean values of $l$, $w$, $\tau$, $v_{j}$, and $a$, which are 3.4$\pm$0.2 Mm, 0.7$\pm$0.2 Mm, 31$\pm$7 s, 220$\pm$10 km s$^{-1}$, and 15$\pm$1 km s$^{-2}$, respectively.

\subsection{Electron Densities and Temperatures} \label{subsec:}
Unfortunately, all of the mini-jets in our study were missed by the $IRIS$ spectrometer slit.
Thus, we cannot directly measure the electron densities ($n_{e}$) of the mini-jets by using the intensity ratio of the \ion{O}{4} 1401\AA\ and 1399 \AA\ line pair.
Here, we provide a rough method for the diagnosis of $n_{e}$.
We found that some places scanned by the slit have similar 1330 \AA\ intensities ($I$) to those of the jets.
Based on the assumption that they may have similar values of $n_{e}$, we first derived the electron densities of the scanned regions from the $IRIS$ spectral data, which are shown by the plus signs in Figure~3(a).
It can be seen that $n_{e}$ increases with the enhancement of $I$ at first and then keeps stable when $I$ exceeds $\sim$2300 DN (DN is data number).
We performed a quadratic-polynomial fitting to the data with $I$ in the range [320, 3500] DN.
The fitting result is indicated by the red curve in Figure~3(a), which seems to fit the data well when $I$ is below 2300 DN.
The relationship between $n_{e}$ and $I$ within this range can be expressed by
\begin{equation}
log(n_{e})=10.0+1.05\times10^{-3}\times I-2.4\times10^{-7}\times I^2
\end{equation}
Then, we calculated $n_{e}$ for each mini-jet according to their individual 1330 \AA\ intensity and Equation~(1) (see the eighth column of Table~1 and Figure~3(b)).
It should be noted that this method only provides a very rough estimate of the density as we assume that the \ion{O}{4} densities are somehow related to \ion{C}{2} emission, which can be invalid for various reasons, e.g., \ion{C}{2} emission can be optically thick, the filling factors of \ion{O}{4} and \ion{C}{2} emission can be different, the plasma seen in \ion{C}{2} and \ion{O}{4} can be unrelated, etc.
The distribution of $n_{e}$ is also displayed by the histogram in Figure~2(f).
Our results show that most electron densities range from 1.1$\pm$0.4 to 3.7$\pm$1.2 $\times$ $10^{10}$ cm$^{-3}$, apart from three values for J18, J20, and J21, namely, 13$\pm$4, 7.9$\pm$2.5, and 10$\pm$3 $\times$ $10^{10}$ cm$^{-3}$, respectively.
The average $n_{e}$ is 2.4$\pm$0.8 $\times$ $10^{10}$ cm$^{-3}$.

The AIA provided a good temporal coverage for the tornado event (see Appendix \ref{sec:appendix} and Figure~\ref{fig9}).
However, due to small scales and/or weak intensities, some mini-jets (e.g. J14, J15, J19, J27, J28, J36, and J37) are hard to observe in the hot EUV lines, especially in AIA 335 \AA\ and 94 \AA.
Most of jets can be detected simultaneously in multiple AIA channels and they evolved roughly identically.
Given the significant response around 10$^{5.5}$ K \citep{martnez11} for the hot AIA EUV wavebands, it is likely that the mini-jets are cool structures.
Similar situations have been discussed by \citet{winebarger13} and \citet{tianh14b}, when they analyzed the temperatures of the inter-moss loops and penumbral bright dots, respectively.
Since the typical method of differential emission measure (DEM) analysis is not sufficiently reliable for determining the temperature due to the poor discrimination at the low temperatures in the AIA channels \citep[e.g.,][]{del11, testa12}, we also applied the EM-Loci technique \citep[e.g.,][]{del02} to determine the likely temperatures of the jets.
The EM-Loci curves of each mini-jet (except for J14, J15, J19, J27, J28, J36, and J37) were obtained by dividing the AIA background-subtracted intensities by the temperature response functions.
J5 and J24 can be observed in the AIA 131, 193, 171, 211, and 335 \AA\ channels.
Their EM-Loci curves are presented in Figure~3(c) and (d), respectively.
As indicated by the black boxes in the panels, there are many crossings of the curves at the low temperatures around 10$^{5.45}$ K, suggesting this is the most likely temperature of J5 and J24.
The centers of the two boxes correspond to log temperatures of 5.46 (J5) and 5.43 (J24), respectively.
Similarly, possible temperatures of the other jets were determined using this method and given in the ninth column of Table~1.
As for J14, J15, J19, J27, J28, J36, and J37, we simply take the mean temperature (10$^{5.42}$ $\approx$ 2.6$\pm$0.1 $\times$ 10$^{5}$ K) of the other jets as theirs.
It is worth pointing out that the mini-jets are most likely multi-thermal. 
The EM-loci method may just help estimate an approximate temperature of the jets.

\subsection{Energetic Characteristics} \label{subsec:}
Considering a model in which the mini-jets are cylinders of fully-ionized ideal gas, we calculated their kinetic and thermal energy densities ($E_{k}$ and $E_{t}$) from the estimated densities and temperatures according to the following equations.
\begin{equation}
E_{k}=\frac{1}{2} \rho v^{2}=\frac{1}{2}n_{e}m_{p}v_{j}^{2}
\end{equation}
\begin{equation}
E_{t}=2n_{e}\frac{3}{2}kT=3n_{e}kT
\end{equation}
Here, $\rho$ is the mass density, $m_{p}$ is the proton mass, and $k$ is the Boltzmann constant.
It should be noted that $v_{j}$ is the jet's apparent velocity in the plane of the sky.
Thus, Equation~(2) only gives lower limits on $E_{k}$.
Our calculations show that $E_{k}$ mainly varies in the range of 1 -- 25 erg cm$^{-3}$ with a mean value of $\sim$9$\pm$3 erg cm$^{-3}$, while $E_{t}$ mostly ranges from 1 to 5 erg cm$^{-3}$ with an average of $\sim$3$\pm$1 erg cm$^{-3}$.
As for some mini-jets, obvious SJI 1330 \AA\ and AIA EUV brightenings can be detected at their footpoints, implying a likely energy release by magnetic reconnection during the jet formation.  
Omitting the other energies, such as gravitational potential energy and radiation energy, we took the sum of $E_{k}$ and $E_{t}$ as the dissipated magnetic energy density $E_{m}$ \citep[][]{priest14}.
Then, we can estimate the dissipated magnetic field strength ($B$) according to the formula
\begin{equation}
E_{m}=E_{k}+E_{t}=\frac{B^{2}}{8\pi}
\end{equation}
Note that $B$ here is not the actual magnetic field in the jets but represents the amount of magnetic field that is converted into accelerating and heating the jet.
The values and distributions of $E_{k}$, $E_{t}$, $E_{m}$, and $B$ are presented in the last few columns of Table~1 and Figure~4, respectively.
Among the 43 mini-jets, J18 is a special one with higher level of energies and field strength, which seems to be associated with its much larger electron density. 
The mean $E_{m}$ and $B$ are 12$\pm$3 erg cm$^{-3}$ and 16$\pm$2 G, respectively.
On average, $E_{k}$ is three or four times larger than $E_{t}$, so that much more magnetic energy was converted into kinetic energy than heat.
Figure~5(a) and (b) separately present the variation of $E_{m}$ and $B$ with $E_{k}$.
Simple linear relationships seem to exist between their logarithms.
According to our fitting, as indicated by the red line in Figure~5(a), the relation between $E_{m}$ and $E_{k}$ is
\begin{equation}
E_{m}=1.86 \ast E_{k}^{0.84}
\end{equation}
The fit for $B$ in Figure~5(b) yields a power law with one half of the index in Equation~(5) because $B$ is proportional to the square root of $E_m$.

\subsection{Characteristic Velocities and Pressures} \label{subsec:}
On the basis of the above results, it is of interest to calculate some typical velocities and pressures associated with the mini-jet activity and analyze their likely relationships.
These parameters include the Alfv\'{e}n speed ($v_{a}$), sound speed ($c_{s}$), gas pressure ($P_{t}$), magnetic pressure ($P_{m}$), and total pressure imposed on the jet ($P_{j}$).
The formulae for the calculations of $v_{a}$ and $c_{s}$ can be expressed as 
\begin{equation}
v_{a}=\frac{B}{\sqrt{4\pi \rho}}
\end{equation}
\begin{equation}
c_{s}=(\frac{2\gamma kT}  {m_{p}} )^{\frac{1}{2}}
\end{equation}
where $\gamma$ is the heat capacity ratio. 
The values of $v_{a}$ and $c_{s}$ for each jet are presented and comparisons made with the jet's apparent velocity ($v_{j}$) in Figure~5(c). 
It can be seen that $v_{a}$ seems to be greater than $v_{j}$, but their differences become smaller as $v_{j}$ increases. 
As for $c_{s}$, it keeps stable with lower values because of its simple form that depends only on the temperature $T$.
Based on the assumption of $E_{m} = E_{k} + E_{t}$, the quantitative relationship between $v_{a}$, $v_{j}$, and $c_{s}$ is derived as
\begin{equation}
v_{a}^2 = v_{j}^2 + 1.8 c_{s}^2
\end{equation}
The respective definitions of $P_{j}$, $P_{m}$, and $P_{t}$ are as follows: 
\begin{equation}
P_{j}=\frac{F}{S}=\frac{Ma}{\pi(w/2)^2}
\end{equation}
\begin{equation}
P_{m}=\frac{B^2}{8\pi}
\end{equation}
\begin{equation}
P_{t}=2n_{e}kT
\end{equation}
where $F$ is the force accelerating the jet and $a$, $M$, $w$, and $S$ are the acceleration, mass, width, and cross sectional area of the jet, respectively.
Figure~5(d) exhibits and compares the three pressure values.
Basically, $P_{j}$ is larger than $P_{m}$ and $P_{t}$.
Their mean values are 19$\pm$6, 12$\pm$3, and 1.8$\pm$0.6 dyn cm$^{-2}$, respectively. 

\subsection{Potential Field Source Surface Extrapolation} \label{subsec:}
A PFSS extrapolation for Carrington rotation 2161 reveals that there existed many loop structures overlying the active region AR 12297, as shown in Figure~6.
In space, these magnetic loops seem to cross the prominence at locations, where the mini-jets occurred.
More interestingly, it can be found that the ejection directions of the jets (indicated by the arrows) are similar to the orientations of the crossed loops.
According to the results from PFSS, the mean background field strength of AR 12297 at the altitude of the tornado is $\sim$12 G, which is roughly compatible with our calculation of the dissipated magnetic field strength $\sim$16 G.
These results suggest that the interaction between the tornado-like prominence and the background field (``external reconnection'') is one of the possible reasons for the production of the mini-jets.
On the other hand, it is well known that a prominence may be contained within a large-scale twisted flux tube \cite[e.g.,][]{Mackay&al2010}. The reconnection of this enveloping field (closely enveloping the prominence) with itself, including the field threading the erupting prominence, (``internal reconnection'') may also produce the mini-jets.
Unfortunately, this event occurred near the solar limb and so nonlinear force-free field extrapolations cannot be employed to help clarify the spatial relationship of the prominence field to its surrounding non-potential field (i.e., the conjectured flux-rope envelope).

\section{Summary and Discussion} \label{sec:summary}
High resolution observations from $IRIS$ SJI and $SDO$ AIA clearly reveal that many single or clustered mini-jets were launched from a tornado-like prominence, which have been rarely reported before.
According to their evolution in $IRIS$ SJI far-UV and AIA EUV channels, the mini-jets are probably small-scale plasma ejections. 
Their average electron density is roughly estimated to be $\sim$2.4 $\times$ $10^{10}$ cm$^{-3}$, similar to that of a typical prominence.
They are likely multi-thermal structures with an approximate mean temperature of $\sim$2.6 $\times$ 10$^{5}$ K.
It has been suggested that some small solar jets can be heated to $\sim$10$^5$ K, such as type II spicules and the transition-region network jets reported by \citet{depontieu07} and \citet{tianh14a}, respectively.
However, chromospheric jets outside or in the penumbra of sunspots studied by \citet{shibata07} and \citet{katsukawa07} seem to possess a much lower temperature ($\sim$10$^4$ K).
The spatial and temporal scales of mini-jets are similar to other small solar jets (see Section~1).
They are mostly a few thousands kilometers long, several hundred kilometers wide, and have a short duration of tens of seconds.
The apparent speed of mini-jets can reach 470 km s$^{-1}$, with most between 100--350 km s$^{-1}$, which seems to be more dynamic than other small jets, especially Type-I spicules \citep{depontieu07} or surges \citep{tianh18} and chromospheric anemone jets \citep{shibata07}, possibly due to the differences in the local plasma environment.

Indeed, the birth place of mini-jets is quite different from other small jets.
They originate from the body of a tornado-like prominence suspended at an altitude of $\sim$30--50 Mm in the corona. 
The other jets including large-scale EUV or X-ray jets reported formerly are basically rooted in the lower solar atmosphere, where the Alfv\'en velocity is typically lower and photospheric flux emergence and cancellation may drive fast reconnection between closed and open fields \citep[e.g.,][]{wangj93, canfield96, pariat10, chenh12} or activate the eruption of a mini-filament \citep[e.g.,][]{moore10, hongj11, sheny12, sterling15}.
The coronal mini-jets presented here have a different origin.
They take place when a tornado prominence has been disturbed and distendeds outwards \citep[see][]{chenh17}.
At this time, magnetic reconnection is likely to occur between the prominence field
and the surrounding field.
The local magnetic energy may be dissipated and converted into heat and kinetic energy by reconnection.
Consequently, the heated prominence material is ejected along the newly-formed fields by enhanced gas pressure and magnetic tension of the reconnected fields.
The schematic diagrams in Figure~7 display such a scenario, suggesting a possible formation mechanism for the mini-jets.  

One must be aware that the prominence may not be in close contact with the background field, but rather be enveloped by a flux rope. This is the case in flux rope models for prominences, which place the prominence material in field line dips under the rope axis (especially for quiescent prominences) or in highly sheared, very flat field around the axis of a so-called hollow-core flux rope (especially for active-region prominences; e.g., \citealt{Bobra&al2008}). Enveloping field may have a much smaller flux content, or be largely absent, in the alternative group of models, which assume that the prominence material resides on long flat field lines in a highly sheared arcade (or, equivalently, in the upper part of a very weakly twisted flux rope). For such relatively simple (smooth) models of prominences in active regions (hollow-core flux rope or highly sheared arcade), the enveloping field is nearly parallel to the field that threads the prominence in the immediate vicinity of the prominence material, and makes a gradual transition to the background field further out. 
The scenario sketched in Figure~\ref{fig7} thus requires that the enveloping field be reconnected away before mini-jets that follow the direction of the background field can form. Such reconnection can indeed occur, especially in the case of confined eruptions, when the background field strongly resists the rising flux rope. A striking example is the confined filament eruption described in \citet{JiH&al2003} and \citet{Alexander&al2006}, which showed heated filament plasma draining back to the solar surface from the top of the halted filament along previously invisible paths. The numerical modeling of the event \citep{Torok&Kliem2005, Hassanin&Kliem2016} demonstrated that the whole flux rope can reconnect with the overlying background field and that the draining paths followed the background field after the reconnection. The new field connections became visible only after the flux threading the filament began to reconnect, so that the filament material traced them. Different from that case, a complete reconnection of the erupting flux does not happen in the event investigated here, since most of the original prominence threads are not destroyed.

Alternatively, considering a possibly high degree of complexity of a tornado prominence's field structure, the coronal mini-jets may be created by many small-scale internal reconnections between nearby threads, which convert magnetic energy into the heating and acceleration of small jets. This may also be implicated in the eruptive instability of a prominence or coronal mass ejection. The threads may either be braided around one another and start reconnecting when the braiding becomes too great or they may each be internally twisted (Figure~8). In both cases, reconnection in one of the threads may start an avalanche of reconnections in the other threads. The reason that the jets are ejected roughly perpendicular to the overall prominence flux rope is that the fibrils are weakly twisted or braided, so that it is the transverse components of the magnetic field in the threads that are reconnected rather than the axial component directed along the flux rope. Reconnection of many small twisted threads and has been modelled numerically by \citet{hood16} and \citet{reid20}, building on earlier numerical MHD models for the formation of many fine-scale currents by kink instability \citep{browning08, hood09}. In practice the structure will be much more complex than indicated in Figure~8, as can be seen in the computations of \citet{hood16}. On the other hand, braiding has been modelled numerically by, for instance, \citet{wilmot10, wilmot11} and \citet{pontin11}. The twisting or braiding of individual threads would naturally be produced by photospheric motions in the photospheric magnetic carpet of the many internal intense flux tubes that produce the magnetic field of a huge prominence flux rope.
The advantage of an explanation in terms of internal reconnection of prominence threads is that it explains in a natural way the fine-scale nature of the mini-jets, their appearance as a cluster, and their direction perpendicular to the prominence.

In our observations, brightenings appeared at the footpoints of some mini-jets and most of the jets were also brightened along their whole lengths, compared to the threads in the swirling prominence. 
It is hard to believe that these brightenings resulted from plasma density enhancements by material accumulations.
In addition, the acceleration of mini-jets can be easily detected (see the online animated version of Figure~\ref{fig1}).
Such observations support a reconnection explanation for mini-jets' formation.
EUV and/or microwave brightenings have been found inside erupting filaments, as reported by \citet{schrijver08} and \citet{huangj19}, which suggest the occurrences of local magnetic energy release by many small-scale internal or external reconnections of a prominence flux rope.
However, no obvious plasma ejections in the form of mini-jets were observed in these events.
\citet{huangz18} found that some jet threads appeared along a large-scale loop in the course of the eruption of a spiral filament.
They found that magnetic reconnections probably occurred at the footpoints of the jets and accelerated them similar to our event.
Recently, \citet{chitta19} reported hot spicules with much lower speed launched from a quiescent turbulent cool prominence, which seem to be generated instead by turbulent motions. 

According to the external reconnection explanation for mini-jets, bi-directional reconnection outflows should be formed along not only the background or enveloping fields but also the tornado fields, as indicated by the red arrows in Figure~7(b).
In several jet cases, such as J41--J43, we indeed observed some bright flows out of the jet footpoints along the prominence's threads.
However, most of the mini-jets were found to be directed almost perpendicular to the prominence axis (likely along the background or enveloping field).
This may be associated with the gas or magnetic pressure difference between the background or enveloping field and tornado field.
The inflating jet plasma tends to move toward the weaker gas or magnetic pressure region (background or enveloping field), as found in MHD simulations of asymmetric magnetic reconnection \citep{cassak07, murphy12}. 
Additionally, any jet component along the prominence threads would be less visible than a component along the background or enveloping field if the threads point more perpendicularly to the sky plane than the latter field. This is quite likely from the geometry of the prominence, which partly drained to foot points behind the limb.

So far, there are very few reports about coronal reconnection mini-jets and so they are worth exploring in more detail in future, in particular with high-resolution observations.
They are associated with active region prominences, especially when activated \citep{chenh17} or even erupting \citep{huangz18}, and so it will be worth determining whether they also take place in erupting quiescent prominences.
In addition, nonlinear force-free \citep[e.g.,][]{mackay06, wiegelmann06, wiegelmann12,  mackay12} or other non-potential \citep[e.g.,][]{zhux17} field extrapolations can help clarify the nature of the tornado magnetic fields and their spatial relationship to the overlying magnetic arcade.
Numerical simulation studies of such jets will also provide us with better understanding of these small-scale plasma ejections.
From a wider point of view, they suggest that solar activities over widely different scales are often coupled together.
Detailed investigations of their association would help a more comprehensive understanding of solar activity.

\acknowledgments
We thank Prof. Hui Tian of Peking University for insightful suggestions and informative discussions.
$IRIS$ is a NASA small explorer mission developed and operated by LMSAL with mission operations executed at NASA Ames Research center and major contributions to downlink communications funded by ESA and the Norwegian Space Centre.
The $SDO$ data are courtesy of NASA, the $SDO$/AIA, and $SDO$/HMI science teams.
This work is supported by NSFC (11790304, 11790301, 11533008, 11941003, 11790300, 41331068, 11673034, 11673035, 11773039, 11973057), the B-type Strategic Priority Program of the Chinese Academy of Sciences, Grant No. XDB41000000 and Key Programs of the Chinese Academy of Sciences (QYZDJ- SSW-SLH050).

\appendix
\section{AIA/SDO images of the mini-jets}
\label{sec:appendix}

Figure~\ref{fig9} displays an online animation of the AIA 094, 131, 193, 171, 211, 304, and 335 \AA\ channels. It runs from 09:10 UT to 10:00 UT, including all of the mini-jets listed in Table~\ref{table1}. 




\clearpage
\begin{deluxetable}{l|cccccc|cc|cccc}
\tabletypesize{\scriptsize} 
\tablecaption{Dynamical and energetic characteristics of the mini-jets\label{table1}}
\tablenum{1}
\tablehead{
\colhead{Jet} & \colhead{Time\tablenotemark{$\star$}} & \colhead{$l$\tablenotemark{*}} &\colhead{$w$\tablenotemark{*}} &\colhead{$v_{j}$\tablenotemark{*}}  & \colhead{$a$\tablenotemark{*}} &\colhead{$\tau$\tablenotemark{*}} & \colhead{$n_{e}$\tablenotemark{*}} & \colhead{$T\tablenotemark{*}$}& \colhead{$E_{k}$\tablenotemark{*}} & \colhead{$E_{t}$\tablenotemark{*}} & \colhead{$E_{m}\tablenotemark{*}$} & \colhead{$B$\tablenotemark{*}} 
\\
\colhead{} & \colhead{(UT)} & \colhead{(10$^3$ km)} &\colhead{(10$^2$ km)} &\colhead{(10$^2$ km s$^{-1}$)} &\colhead{(km s$^{-2}$)} &\colhead{(s)} &  \colhead{(10$^{10}$ cm$^{-3}$)} & \colhead{(10$^5$ K)}   &\colhead{(erg cm$^{-3}$)} &\colhead{(erg cm$^{-3}$)} &\colhead{(erg cm$^{-3}$)} &\colhead{(Gauss)}
 }
\startdata
J1   & 09:17:25 & 3.0$\pm$0.2   & 6$\pm$2  &  1.7$\pm$0.1  & 9$\pm$1  & 28$\pm$7 & 1.1$\pm$0.4 & 2.5$\pm$0.1  & 2.6$\pm$0.9  &  1.2$\pm$0.4  & 3.8$\pm$1.0   &  10$\pm$1 \\
J2   & 09:20:50 & 3.0$\pm$0.2  & 5$\pm$2  &  1.6$\pm$0.1  & 9$\pm$1  & 28$\pm$7  & 1.7$\pm$0.5 & 2.6$\pm$0.1  & 3.9$\pm$1.3     &  1.9$\pm$0.6  &  5.8$\pm$1.4 &   12$\pm$2 \\
J3   & 09:21:09 & 5.1$\pm$0.2  & 5$\pm$2  &  2.7$\pm$0.1  & 14$\pm$1  &  28$\pm$7 & 2.1$\pm$0.7 & 2.7$\pm$0.1  & 13$\pm$4  &  2.4$\pm$0.8  & 15$\pm$4 & 20$\pm$3 \\  
J4   & 09:21:09 & 5.5$\pm$0.2  & 7$\pm$2  &  2.0$\pm$0.1  & 7$\pm$1  & 38$\pm$7  & 2.4$\pm$0.7 & 2.6$\pm$0.1  & 7.6$\pm$2.5  & 2.6$\pm$0.8 &  10$\pm$3 & 16$\pm$2 \\  
J5   & 09:21:09 & 3.1$\pm$0.2  & 6$\pm$2  &  1.1$\pm$0.1  & 4$\pm$1  &  38$\pm$7  & 3.2$\pm$1.0 & 2.9$\pm$0.1  & 3.3$\pm$1.2  & 3.9$\pm$1.2  & 7.2$\pm$1.7   &   14$\pm$2 \\
J6   & 09:21:09 & 2.8$\pm$0.2  & 6$\pm$2  & 1.0$\pm$0.1  & 4$\pm$1 &  38$\pm$7 &  3.2$\pm$1.0 & 2.8$\pm$0.1  & 2.7$\pm$1.0  & 3.8$\pm$1.2  & 6.5$\pm$1.6  & 13$\pm$2 \\ 
J7   & 09:21:46 & 5.1$\pm$0.2  &  7$\pm$2  & 1.8$\pm$0.1  & 7$\pm$1  & 46$\pm$7 & 1.5$\pm$0.5 & 2.6$\pm$0.1  & 4.2$\pm$1.4  & 1.6$\pm$0.5  & 5.8$\pm$1.5  &    12$\pm$2  \\   
J8   & 09:21:55 & 4.6$\pm$0.2  & 7$\pm$2   &  2.4$\pm$0.1  & 13$\pm$1  & 37$\pm$7 & 2.4$\pm$0.8 & 2.6$\pm$0.1  & 12$\pm$4  & 2.6$\pm$0.8 &  15$\pm$4  & 19$\pm$3 \\ 
J9   & 09:21:55 & 5.7$\pm$0.2  & 7$\pm$2 &   2.0$\pm$0.1   & 7$\pm$1  & 46$\pm$7  & 2.8$\pm$0.9  & 2.7$\pm$0.1 & 9.7$\pm$3.2    & 3.1$\pm$1.0 &  13$\pm$3  & 18$\pm$2 \\ 
J10 & 09:22:14 & 3.6$\pm$0.2  & 8$\pm$2  &   2.0$\pm$0.1 & 11$\pm$1  & 37$\pm$7 & 3.1$\pm$1.0 & 2.8$\pm$0.1  & 11$\pm$4  & 3.6$\pm$1.1 & 14$\pm$4  &   19$\pm$2 \\  
J11 & 09:22:14 & 2.6$\pm$0.2  &  7$\pm$2   &  2.9$\pm$0.1  &  32$\pm$1   &  28$\pm$7 & 3.4$\pm$1.0 &  2.6$\pm$0.1  &  23$\pm$7    & 3.7$\pm$1.2  &  27$\pm$7 &  26$\pm$4 \\   
J12 & 09:22:32 & 1.7$\pm$0.2   &  3$\pm$2  & 1.9$\pm$0.1   &  21$\pm$1   &  19$\pm$7 & 1.5$\pm$0.5  &  2.7$\pm$0.1  & 4.3$\pm$1.4   & 1.6$\pm$0.5  & 5.9$\pm$1.5   &  12$\pm$2 \\   
J13 & 09:26:25 & 2.7$\pm$0.2   &  6$\pm$2  & 1.4$\pm$0.1   &  8$\pm$1  &   28$\pm$7 & 1.1$\pm$0.3 & 2.8$\pm$0.1 &  1.9$\pm$0.6   &   1.2$\pm$0.4   &  3.1$\pm$0.8   &   9$\pm$1 \\    
J14 & 09:26:53 & 1.9$\pm$0.2  &   5$\pm$2  &  1.9$\pm$0.1   &  19$\pm$1   &  19$\pm$7 & 1.1$\pm$0.3 & 2.6$\pm$0.1\tablenotemark{$\diamond$}  &  3.3$\pm$1.1   & 1.2$\pm$0.4   &   4.5$\pm$1.1   &  11$\pm$1 \\     
J15 & 09:26:53 & 3.1$\pm$0.2   &  6$\pm$2  &  1.6$\pm$0.1   &  8$\pm$1   &  28$\pm$7  & 1.2$\pm$0.4  & 2.6$\pm$0.1\tablenotemark{$\diamond$}  &  2.7$\pm$0.9   & 1.3$\pm$0.4   &  4.0$\pm$1.0  & 10$\pm$1 \\   
J16  & 09:28:45 & 1.6$\pm$0.2   &  6$\pm$2  &  1.8$\pm$0.1   & 20$\pm$1   &  18$\pm$7 & 1.4$\pm$0.4  &  2.7$\pm$0.1  &  3.8$\pm$1.3   &  1.6$\pm$0.5   &   5.4$\pm$1.4  &    12$\pm$1 \\     
J17 & 09:28:45 & 3.1$\pm$0.2    &  5$\pm$2  &   1.1$\pm$0.1   &  4$\pm$1   &  28$\pm$7  & 1.6$\pm$0.5 &   2.6$\pm$0.1  & 1.7$\pm$0.6  &  1.7$\pm$0.6   &   3.4$\pm$0.8   &    9$\pm$1 \\    
J18 & 09:29:50 & 4.3$\pm$0.2  &   6$\pm$2   &  2.3$\pm$0.1  &  12$\pm$1   &  28$\pm$7 & 13$\pm$4 &  2.7$\pm$0.1  &   57$\pm$19   &  15$\pm$5   & 72$\pm$19 &  43$\pm$6 \\   
J19 & 09:30:28& 3.3$\pm$0.2  &   6$\pm$2   &  3.3$\pm$0.1   &  33$\pm$1   &  19$\pm$7 & 1.1$\pm$0.4 & 2.6$\pm$0.1\tablenotemark{$\diamond$}    & 10$\pm$3   &  1.3$\pm$0.4   & 12$\pm$3  &   17$\pm$2 \\     
J20 & 09:30:55 & 2.6$\pm$0.2   &  6$\pm$2  &   1.4$\pm$0.1   &  8$\pm$1  &   37$\pm$7 & 7.9$\pm$2.5  & 2.8$\pm$0.1  &  14$\pm$5  & 9.3$\pm$2.9 & 23$\pm$6 &   24$\pm$3 \\      
J21 & 09:31:05 & 2.4$\pm$0.2   &  6$\pm$2  &   1.3$\pm$0.1   &  7$\pm$1   &  28$\pm$7 & 10$\pm$3 & 2.7$\pm$0.1   &  14$\pm$5  &  11$\pm$4   &  25$\pm$6 &  25$\pm$3 \\      
J22 & 09:31:05 & 5.3$\pm$0.2  &   7$\pm$2   &  2.8$\pm$0.1   &  15$\pm$1   &  28$\pm$7 & 1.2$\pm$0.4 & 2.6$\pm$0.1  &  7.6$\pm$2.5   & 1.3$\pm$0.4  & 8.9$\pm$2.5  &   15$\pm$2 \\       
J23 & 09:31:14 & 7.6$\pm$0.2   &  6$\pm$2   &  2.7$\pm$0.1   &  10$\pm$1   &  47$\pm$7 & 1.4$\pm$0.4  & 2.7$\pm$0.1   & 8.5$\pm$2.7   &  1.5$\pm$0.5  & 10$\pm$3 &   16$\pm$2 \\     
J24 & 09:31:14& 6.8$\pm$0.2  &   8$\pm$2   &   3.8$\pm$0.1   &  21$\pm$1  &   37$\pm$7 & 1.2$\pm$0.4 & 2.7$\pm$0.1 &   14$\pm$5 &  1.3$\pm$0.4   &  16$\pm$5 &   20$\pm$3  \\      
J25 & 09:31:14& 5.3$\pm$0.2   &  8$\pm$2   &  2.9$\pm$0.1   &  16$\pm$1  &   37$\pm$7 & 1.4$\pm$0.4 & 2.7$\pm$0.1   & 10$\pm$3  &  1.6$\pm$0.5  &  12$\pm$3 &  17$\pm$2  \\      
J26 & 09:31:14& 3.6$\pm$0.2  &   8$\pm$2   &  2.0$\pm$0.1   &  11$\pm$1  &   37$\pm$7 & 1.4$\pm$0.4 & 2.6$\pm$0.1  &  4.7$\pm$1.6  &  1.6$\pm$0.5  &   6.3$\pm$1.6  &   13$\pm$2  \\      
J27 & 09:32:01& 4.7$\pm$0.2   &  6$\pm$2   &  4.7$\pm$0.1  &   47$\pm$1   &  19$\pm$7 & 1.1$\pm$0.4 & 2.6$\pm$0.1\tablenotemark{$\diamond$}    &  21$\pm$7  &  1.2$\pm$0.4   &   22$\pm$7    &  24$\pm$4  \\      
J28 & 09:32:01& 4.2$\pm$0.2   &  5$\pm$2   &  4.2$\pm$0.1  &   42$\pm$1   &  19$\pm$7 & 1.1$\pm$0.4 & 2.6$\pm$0.1\tablenotemark{$\diamond$}    &  17$\pm$5  &  1.3$\pm$0.4  &   18$\pm$5   &   21$\pm$3  \\     
J29 & 09:36:03& 3.7$\pm$0.2   &  5$\pm$2   &  2.1$\pm$0.1  &   11$\pm$1  &   46$\pm$7 & 1.6$\pm$0.5 & 2.8$\pm$0.1   &  5.8$\pm$1.9  &  1.9$\pm$0.6  &    7.7$\pm$2.0    &    14$\pm$2  \\      
J30 & 09:36:03& 2.8$\pm$0.2  &   5$\pm$2   &  3.2$\pm$0.1  &   35$\pm$1  &   37$\pm$7 & 1.2$\pm$0.4 & 2.5$\pm$0.1  &   10$\pm$3  &  1.3$\pm$0.4  &   11$\pm$3 &     17$\pm$2  \\     
J31 & 09:37:17&  3.3$\pm$0.2   &  6$\pm$2  &  1.8$\pm$0.1    &  10$\pm$1     &   37$\pm$7 &  2.9$\pm$0.9  &  2.7$\pm$0.1   &  8.0$\pm$2.7   &  3.2$\pm$1.0 & 11$\pm$3   &   17$\pm$2 \\    
J32 & 09:38:04&  3.6$\pm$0.2   &   5$\pm$2  & 2.0$\pm$0.1    &   11$\pm$1   &    37$\pm$7 & 3.7$\pm$1.2 &  2.7$\pm$0.1   &   12$\pm$4    &  4.1$\pm$1.3 & 16$\pm$4 &   20$\pm$3 \\   
J33 & 09:38:23&  5.7$\pm$0.2   &   12$\pm$2  &  3.0$\pm$0.1    &  16$\pm$1   &    47$\pm$7 & 1.6$\pm$0.5  &  2.6$\pm$0.1   &   12$\pm$4    & 1.8$\pm$0.6 & 14$\pm$4 &   19$\pm$3   \\   
J34 & 09:39:19&  2.0$\pm$0.2   &   4$\pm$2  &    2.2$\pm$0.1    &   25$\pm$1   &    18$\pm$7 & 1.5$\pm$0.5 &  2.4$\pm$0.1   &  6.2$\pm$2.0  &   1.5$\pm$0.5  &  7.7$\pm$2.1   &   14$\pm$2 \\    
J35 & 09:39:19&  4.1$\pm$0.2    &  3$\pm$2    &   2.3$\pm$0.1   &    13$\pm$1   &    27$\pm$7 &  1.8$\pm$0.6 &  2.6$\pm$0.1   &   7.5$\pm$2.5  & 1.9$\pm$0.6   &   9.4$\pm$2.5   &   15$\pm$2  \\   
J36  &09:26:07 &  2.0$\pm$0.2    &   4$\pm$2    &  2.0$\pm$0.1   &    20$\pm$1   &    19$\pm$7 & 1.1$\pm$0.3  &  2.6$\pm$0.1\tablenotemark{$\diamond$}     &   3.6$\pm$1.2   &   1.2$\pm$0.4   &    4.8$\pm$1.3  &     11$\pm$1   \\   
J37  &09:26:07 &  2.7$\pm$0.2    &   4$\pm$2   &   3.1$\pm$0.1    &    34$\pm$1   &    18$\pm$7 &  1.1$\pm$0.3  & 2.6$\pm$0.1\tablenotemark{$\diamond$}     & 8.6$\pm$2.8   &    1.2$\pm$0.4   &   9.8$\pm$2.8    &    16$\pm$2  \\  
J38  &09:26:16 &  5.8$\pm$0.2   &   5$\pm$2   &   3.2$\pm$0.1   &    18$\pm$1   &   28$\pm$7 & 1.1$\pm$0.3 &  2.6$\pm$0.1    & 9.4$\pm$3.0  & 1.2$\pm$0.4   &     11$\pm$3   &   16$\pm$2  \\    
J39  &09:24:24 &  2.1$\pm$0.2    &   6$\pm$2    &   1.1$\pm$0.1    &   6$\pm$1   &    28$\pm$7 & 1.2$\pm$0.4  &  2.5$\pm$0.1    &  1.2$\pm$0.5  & 1.3$\pm$0.4   & 2.5$\pm$0.6 &     8$\pm$1  \\  
J40  &09:24:52 &  3.9$\pm$0.2   &    13$\pm$2   &   1.4$\pm$0.1    &   5$\pm$1   &    37$\pm$7 & 1.9$\pm$0.6  &  2.5$\pm$0.1   &   3.0$\pm$1.0   &  1.9$\pm$0.6    &   4.9$\pm$1.2   &   11$\pm$1   \\   
J41  &09:25:29 &  3.3$\pm$0.2    &   11$\pm$2   &    1.8$\pm$0.1    &   9$\pm$1   &    37$\pm$7 & 1.3$\pm$0.4  &  2.7$\pm$0.1    &  3.5$\pm$1.2   & 1.5$\pm$0.5  &    5.0$\pm$1.2  &   11$\pm$1  \\   
J42  &09:25:20 &  4.7$\pm$0.2   &    19$\pm$2   &    2.6$\pm$0.1    &   14$\pm$1    &   37$\pm$7 & 2.9$\pm$0.9 &  2.5$\pm$0.1   &  17$\pm$5   &   3.0$\pm$1.0  &   20$\pm$5   &   22$\pm$3  \\    
J43  &09:25:29 &  4.4$\pm$0.2   &    13$\pm$2    &   2.3$\pm$0.1    &   12$\pm$1   &    28$\pm$7 & 1.1$\pm$0.4  &  2.6$\pm$0.1   &  5.0$\pm$1.6    &   1.2$\pm$0.4  &   6.2$\pm$1.7   &  12$\pm$2   \\  
\cline{1-13}
Mean:  & &  3.4$\pm$0.2   &    7$\pm$2    &   2.2$\pm$0.1    &   15$\pm$1   &    31$\pm$7 & 2.4$\pm$0.8 & 2.6$\pm$0.1    &    9.3$\pm$3.1    &  2.7$\pm$0.8    &   12$\pm$3  &  16$\pm$2  \\
\enddata
\tablenotetext{ }{\textbf{Notes.}}
\tablenotetext{\star}{Means the time when the mini-jet first appeared in the 1330 \AA\ SJI image.}
\tablenotetext{*}{Denote the mini-jet's projected length ($l$), width ($w$), velocity ($v_{j}$), acceleration ($a$), lifetime ($\tau$), temperature ($T$), electron density ($n_{e}$), kinetic energy density ($E_{k}$), thermal energy density ($E_{t}$), dissipated magnetic energy density ($E_{m}=E_{k}+E_{t}$), and magnetic field strength ($B$), respectively.}
\tablenotetext{\diamond}{Due to being undetectable in the AIA EUV lines, we simply take the mean temperature of the other jets as the temperatures of J14, J15, J19, J27, J28, J36, and J37.}
\end{deluxetable}
\begin{figure}
\epsscale{0.95}
\plotone{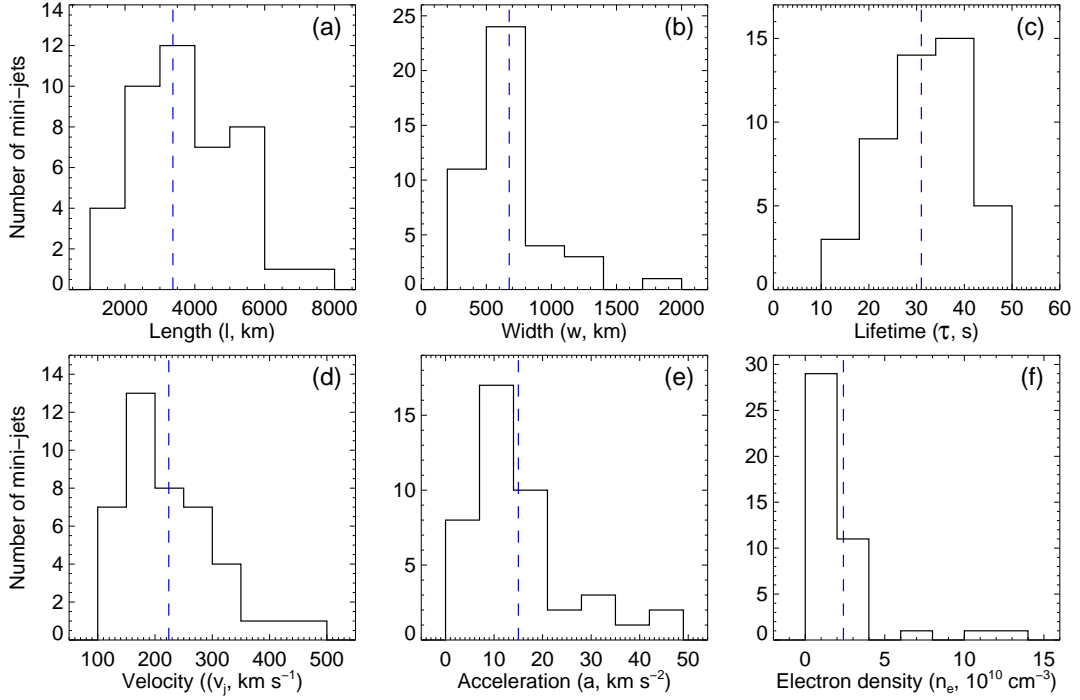}
\caption{(a)--(f) Distributions of the length, width, lifetime, velocity, acceleration, and electron density for the mini-jets. The dashed lines indicate the respective mean values.
\label{fig2}}
\end{figure}

\begin{figure}
\epsscale{0.7}
\plotone{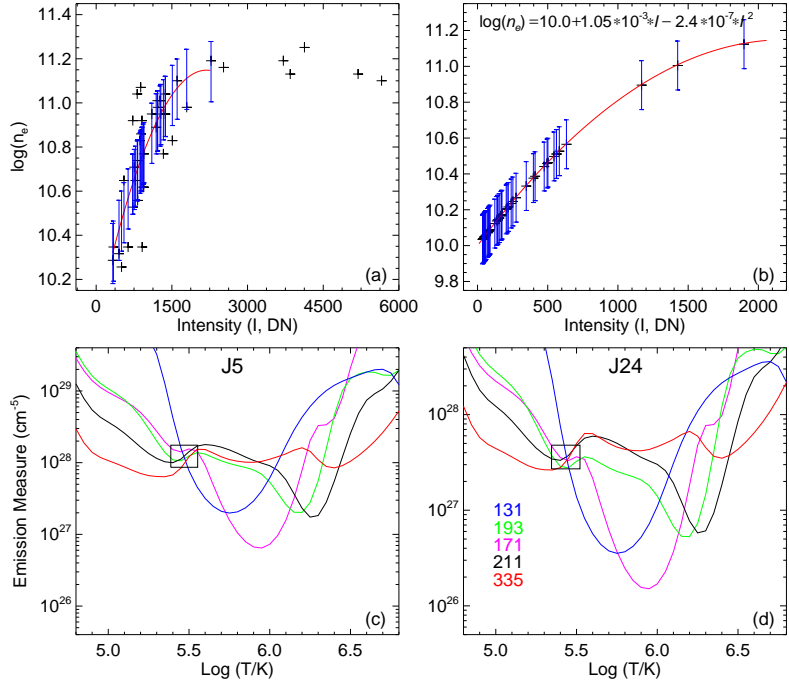}
\caption{(a) The relationship between the 1330 \AA\ intensity ($I$) and the electron density ($n_{e}$) derived from the intensity ratio of the \ion{O}{4} 1401\AA\ and 1399 \AA\ line pair; The red curve is the quadratic-polynomial fitting result with 1$\sigma$ error bar to the data with intensity in the range [320, 3500]; (b) log($n_{e}$) with 1$\sigma$ error bar of mini-jets derived from the fitting curve in panel (a).
(c)--(d) The EM-loci curves for J5 and J24, respectively. 
The black boxes show the regions with many crossings of the EM-loci curves.
\label{fig3}}
\end{figure}

\begin{figure}
\epsscale{0.75}
\plotone{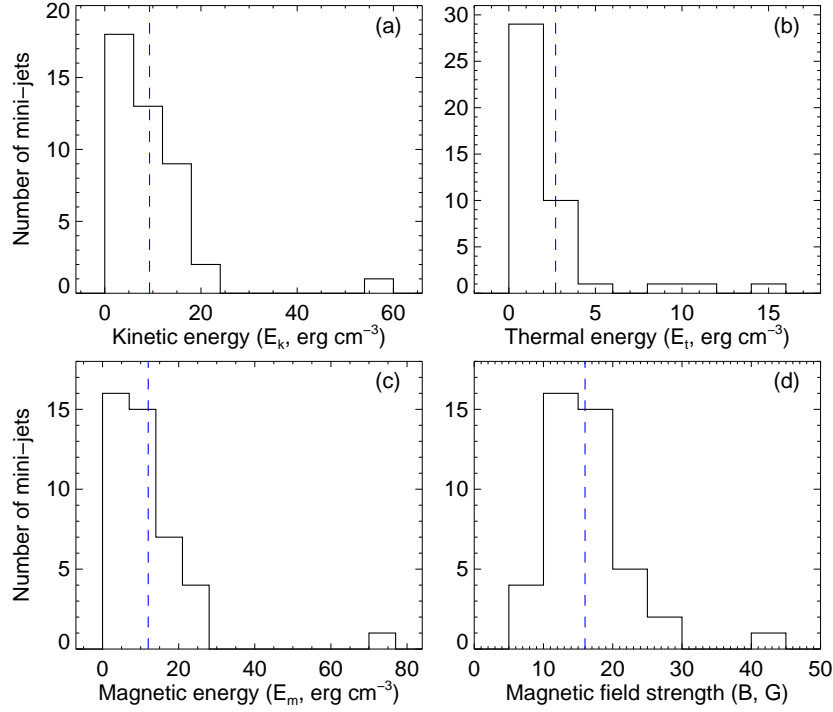}
\caption{(a)--(f) Distributions of the kinetic energy, thermal energy, magnetic energy, and magnetic field strength for the mini-jets. The dashed lines correspond to the respective mean values.
\label{fig4}}
\end{figure}

\begin{figure}
\epsscale{0.75}
\plotone{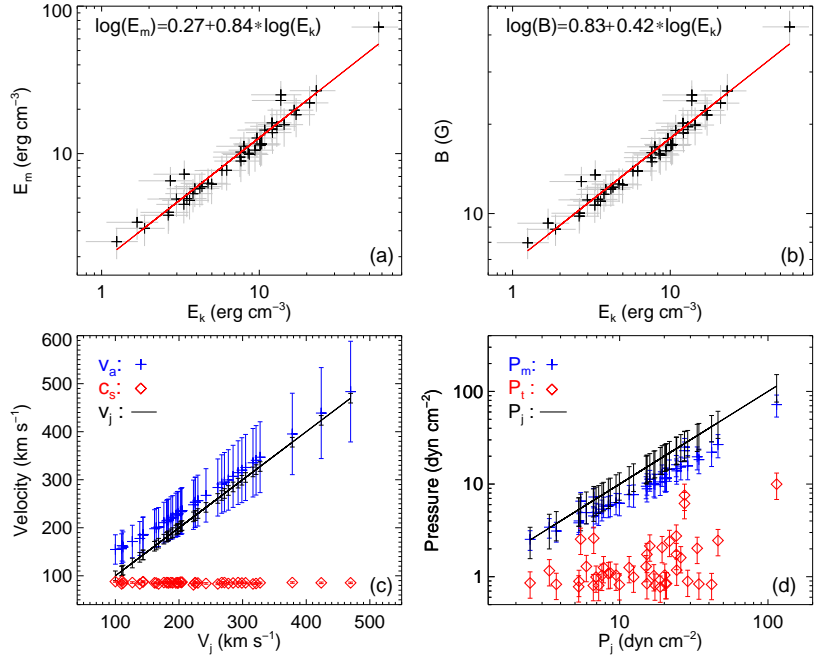}
\caption{The variation relations between $E_{k}$ and $E_{m}$ (a), and between $E_{k}$ and $B$ (b).
The red lines show the linear-fitting results of their logarithms.
The gray lines in panels (a) and (b) are the 1$\sigma$ uncertainties of $E_{k}$, $E_{m}$, and $B$.
Panel (c) shows the comparisons between the local Alfv\'{e}n speed ($v_{a}$), sound speed ($c_{s}$) and the jet's apparent velocity ($v_{j}$) with their 1$\sigma$ error bars.
Panel (d) presents the magnetic pressure ($P_{m}$), gas pressure ($P_{t}$), and total pressure imposed on the jet ($P_{j}$) with their 1$\sigma$ error bars.
\label{fig5}}
\end{figure}

\begin{figure}
\epsscale{0.7}
\plotone{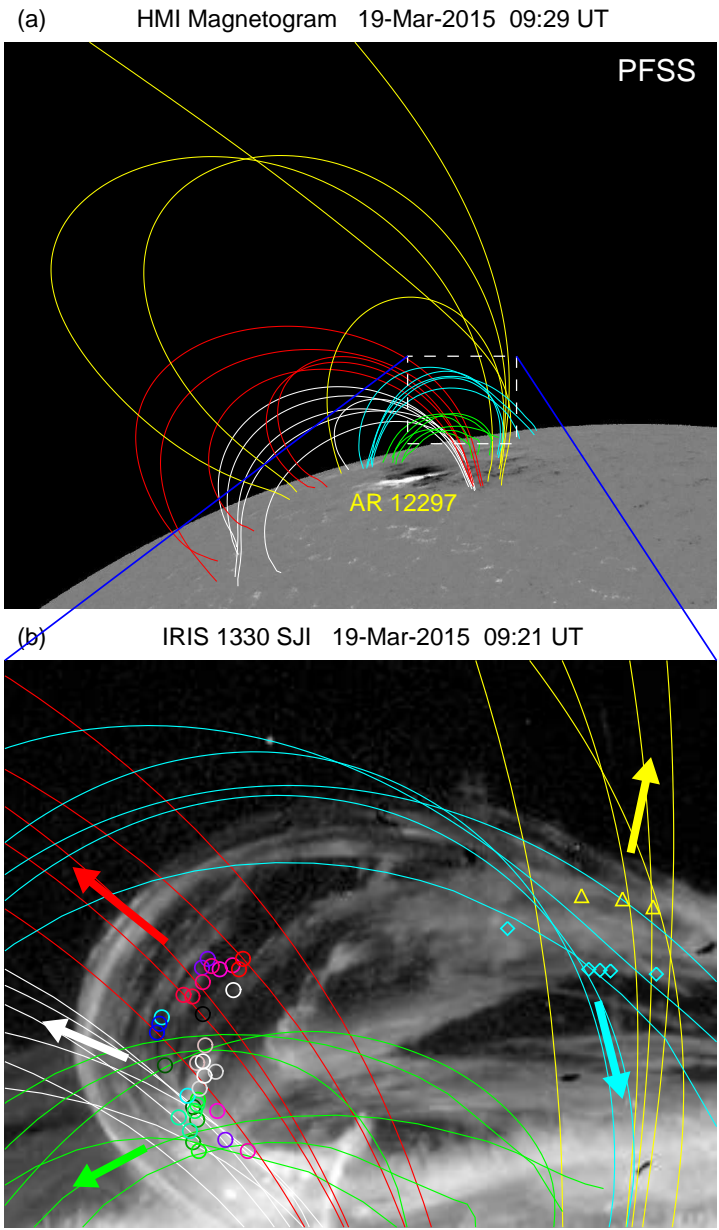}
\caption{Magnetic field lines from a corresponding PFSS extrapolation are overlaid on the HMI magnetogram (a) and IRIS 1330 \AA\ SJI image (b).
The field lines of different colors indicate the different loop structures overlying the active region AR 12297.
The box in panel (a) represent the FOV of panel (b).
Panel (b) is the same image as Figure~1(a) but overlaid with the extrapolated field lines.
\label{fig6}}
\end{figure}

\begin{figure}
\epsscale{1}
\plotone{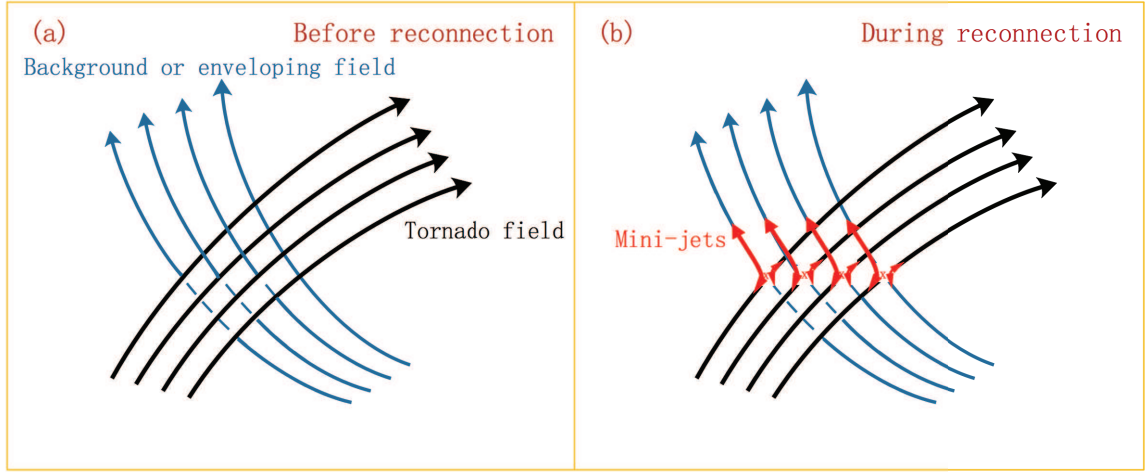}
\caption{The formation of mini-jets by reconnection between the background or enveloping field and the tornado field.
The ``X'' symbols denote the spots where the magnetic reconnections take place between the fields of tornado and background.
\label{fig7}}
\end{figure}

\begin{figure}
\epsscale{1}
\plotone{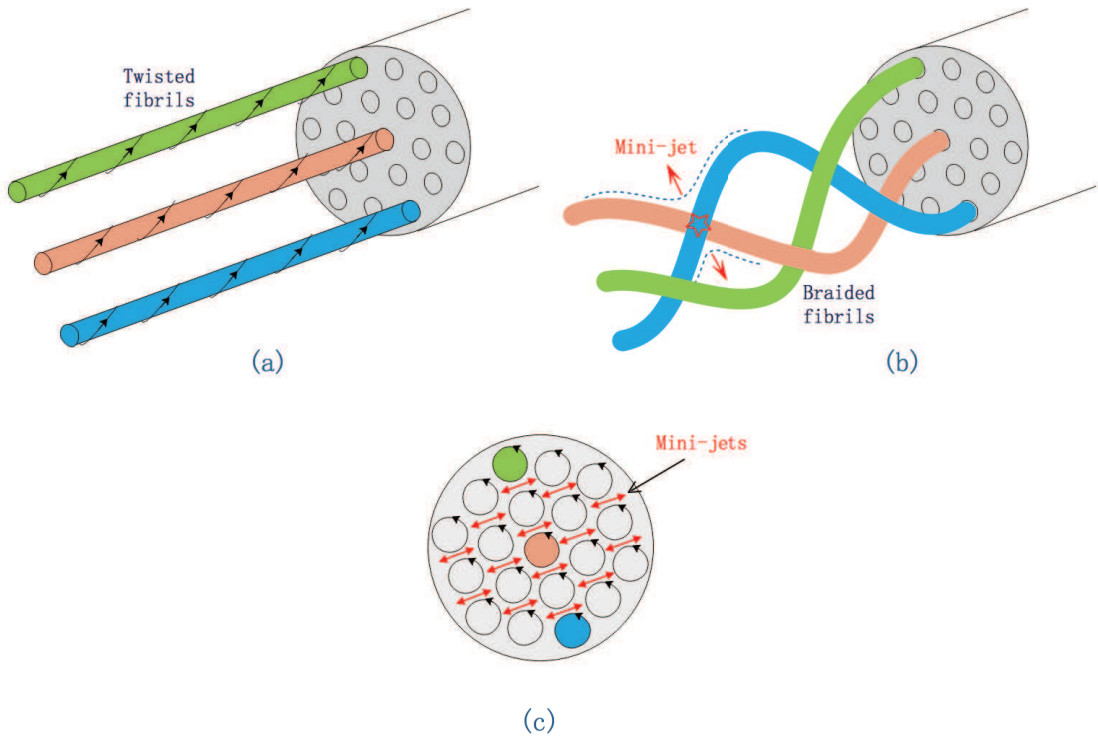}
\caption{Schematics of the formation of mini-jets by internal reconnection between (a) twisted or (b) braided fibrils that make up the magnetic flux rope of the prominence tornado. (a) and (b) indicate the circular cross-section of the flux rope together with the cross-sections of the magnetic fibrils, with three of the fibrils being indicated. (c) The cross-sections of the large-scale flux rope and fibrils, indicating reconnection of the transverse magnetic field of the fibrils and the production of jets (solid-headed arrows) in directions perpendicular to the flux rope.
\label{fig8}}
\end{figure}

\begin{figure}
\epsscale{1}
\plotone{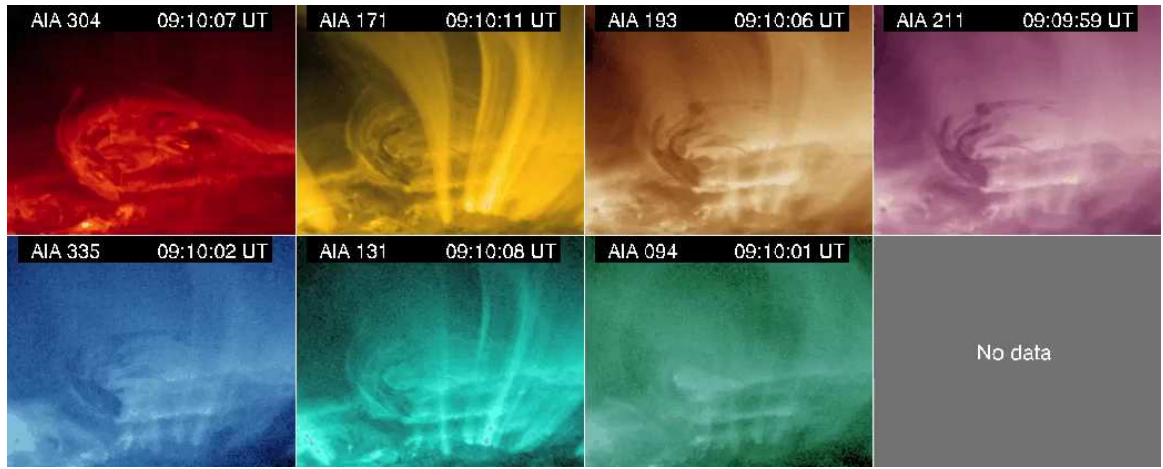}
\caption{Temporal coverage for the tornado event from AIA/SDO. The online animation of the AIA 094, 131, 193, 171, 211, 304, and 335 \AA\ channels runs from 09:10 UT to 10:00 UT, including all of the mini-jets listed in Table~\ref{table1}. 
\label{fig9}}
\end{figure}

\end{CJK*}

\end{document}